\documentclass[seceqn]{elsart}

\usepackage[dvips]{graphics}
\usepackage{graphicx}
\usepackage{hyperref}
\usepackage{amsmath,amssymb}
\usepackage{amsfonts}
\usepackage{morefloats}

\newcounter{bla}

\newcommand{\rd}{\mathrm{d}}

\def\beq{\begin{equation}}
\def\eeq{\end{equation}}
\def\bsp#1\esp{\begin{split}#1\end{split}}

\def \sha{{\,\amalg\hskip -3.8pt\amalg\,}}
\def \uplus{\sha}
\newcommand{\li}[1]{\mathrm{Li}_{#1}}

\newcommand{\cS}{\begin{cal}S\end{cal}}
\newcommand{\cB}{\begin{cal}B\end{cal}}
\newcommand{\cH}{\begin{cal}H\end{cal}}

\renewcommand{\ln}{\log}

\begin{document}

\begin{frontmatter}

\begin{flushright}
IPPP/11/36, DCPT/11/72
\end{flushright}

\title{CHAPLIN - Complex Harmonic Polylogarithms in Fortran}
\author[a]{Stephan Buehler},
\author[b]{Claude Duhr},

\address[a]{Institut f\"ur theoretische Physik, ETH Z\"urich,\\ Wolfgang-Paulistr. 27, CH-8093, Switzerland\\ Email: buehler@itp.phys.ethz.ch}
\address[b]{Institute for Particle Physics Phenomenology, University of Durham,\\ Durham, DH1 3LE, United Kingdom,\\ Email: claude.duhr@durham.ac.uk}

\begin{abstract}
 We present a new Fortran library to evaluate all harmonic polylogarithms up to weight four numerically for any complex argument. The algorithm is based on a reduction of harmonic polylogarithms up to weight four to a minimal set of basis functions that are computed numerically using series expansions allowing for fast and reliable numerical results.

\begin{flushleft}
\end{flushleft}


\end{abstract}

\end{frontmatter}

\newpage


{\bf PROGRAM SUMMARY}

\begin{small}
\noindent
{\em Manuscript Title:} CHAPLIN - Complex Harmonic Polylogarithms in Fortran\\
{\em Authors:}  Stephan Buehler, Claude Duhr\\
{\em Program Title:} Chaplin\\
{\em Journal Reference:}\\
{\em Catalogue identifier:}\\
{\em Licensing provisions:}\\
{\em Programming language:} Fortran 77\\
{\em Computer:} Computing systems on which Fortran 77 compilers are available.\\
{\em Operating system:} Operating systems on which Fortran 77 compilers are available.
{\em Classification:} 11.1 General, High Energy Physics and Computing\\
{\em Nature of problem:} Numerical evaluation of harmonic polylogarithms.\\
{\em Solution method:} Inside the unit circle: series expansion. Outside the unit circle: inversion relations.\\
{\em Restrictions:} Only harmonic polylogarithms up to weight four are supported.\\
{\em Unusual features:} Allows to evaluate HPL's numerically for any point in the complex plane.\\
{\em Running time:} Depending on the weight vector and argument of the HPL, between $0.2$ and $400$ $\mu s$.\\
   \\

\end{small}

\newpage
\section{Introduction}
Feynman integrals in perturbative quantum field theory are generically expressed in terms of the classical polylogarithm functions $\li{n}(z)$ and the Nielsen polylogarithms $S_{n,p}(z)$~\cite{Nielsen}. In the late nineties, it was realized that these classes of functions are too restricted when going beyond one-loop level in the perturbative expansion, where new functions appear that can no longer be expressed in terms of the classical polylogarithm functions. While a completely generic generalization of polylogarithms has been studied in the mathematical literature (going under the name of multiple polylogarithms~\cite{Goncharov:1998, Goncharov:2001}), it is mostly only a specific subset of multiple polylogarithms, the so-called harmonic polylogarithms~\cite{Remiddi:1999ew} and their two-dimensional and cyclotomic generalizations~\cite{Gehrmann:2000zt, Ablinger:2011te}, that make their appearance in the theoretical predictions of physical quantities beyond leading-order. 

In this paper we concentrate exclusively on harmonic polylogarithms (HPL's) up to weight four, which since their introduction have found many applications in computations up to two-loop order in the perturbative expansion, \emph{e.g.},~\cite{Vermaseren:2005qc,   Moch:2004xu,   Vogt:2004mw,    Moch:2004pa,    Bonciani:2003hc,  Bernreuther:2004ih,   Bernreuther:2004th, Bernreuther:2005rw, Mastrolia:2003yz,  Bonciani:2004qt,  Bonciani:2004gi,  Czakon:2004wm,  Bern:2006ew,  Heinrich:2004iq,  Smirnov:2001cm,     Bork:2010wf,   DelDuca:2010zg,  Henn:2010ir,   Aglietti:2006tp,   Aglietti:2004ki,  Aglietti:2004nj,  Gehrmann:2001ck,  Anastasiou:2006hc}. 
In order to confront the theoretical next-to-next-to-leading order (NNLO) predictions to experiment, it is mandatory to be able to evaluate HPL's numerically in a fast and accurate way. The requirements to such a numerical code are twofold: first, the evaluation should be fast, because the use of NNLO matrix elements in Monte Carlo integration codes may require thousands, if not millions, of function calls. Second, it is desirable to be able to compute HPL's for arbitrary complex arguments, which appear for example when the complex mass scheme is employed or for certain kinematic configurations in loop calculations involving massive particles \cite{Bernreuther:2005rw,Anastasiou:2006hc}. In the last decade, various codes have been developed to evaluate HPL's numerically. While the code {\tt hplog}~\cite{Gehrmann:2001pz}, written in {\tt Fortran}, is restricted to the evaluation of HPL's up to weight four and for real values of the arguments, the code {\sc HPL} ({\sc Mathematica})\cite{Maitre:2005uu, Maitre:2007kp} and the implementation of the harmonic polylogarithms into the GiNaC framework ({\tt C++})~\cite{Vollinga:2004sn} are generic and allow to evaluate in principle any harmonic polylogarithm with arbitrary precision for any complex argument.

The focus of this paper is the {\sc Chaplin} (Complex HArmonic PolyLogarithms In fortraN) library, a new {\tt Fortran} code that allows to evaluate numerically all harmonic polylogarithms up to weight four for arbitrary complex arguments. While {\sc Chaplin} is similar in spirit to the aforementioned codes, its main advantages lie in speed, through the use of {\tt Fortran} as a programming language, and in its capability to compute HPL's numerically for any point in the complex plane. {\sc Chaplin} reduces each of the 120 HPL's up to weight four to a set of 32 basis functions~\cite{Gangl:2011}, which are entirely expressed through only two new functions of weight four besides the classical polylogarithms. The basis functions are then mapped to the interior of the unit circle, where they are computed numerically using suitably chosen series expansions that allow to obtain a fast numerical convergence.

The paper is organized as follows: In Section~\ref{sec:hpl_review} we give a short review of harmonic polylogarithms and of their main algebraic and analytic properties. In Section~\ref{reductionchapter} we review the reduction of all HPL's up to weight four to the set of basis functions introduced in Ref.~\cite{Gangl:2011}. The series expansions used by {\sc Chaplin} to compute the basis functions numerically are derived in Section~\ref{sec:series}, while the {\sc Chaplin} library itself, together with comparisons to {\tt hplog}, HPL and GiNaC, is presented in Section~\ref{sec:chaplin}.

\section{Short review of harmonic polylogarithms}
\label{sec:hpl_review}
In this section we give a short review of harmonic polylogarithms (HPL's), as they are at the heart of the {\sc Chaplin} library. HPL's are defined recursively via the iterated integrals~\cite{Remiddi:1999ew}
\beq
\label{eq:hpldef}
H(a_1,\ldots,a_n; z) \, = \, \int_0^z \rd t\,f(a_1;t) \, H(a_2,\ldots,a_n; t)\, ,
\eeq
where $a_i\in\{-1,0,1\}$ and
\beq\label{eq:faz}
f(-1;z) = {1\over 1+z}\,,\quad f(0,z) = {1\over z}\,,\quad f(1;z) = {1\over 1-z}\,,
\eeq
and where we defined $H(;z) = 1$. If all the $a_i$ are simultaneously zero, the integral~\eqref{eq:hpldef} is divergent, and so in this case we use the definition,
\beq\label{eq:H000}
H(\vec 0_n;z) = {1\over n!}\,\log^n z\,,
\eeq
where we used the obvious vector notation $\vec a_n = (\underbrace{a,\ldots, a}_{n\textrm{ times}})$. Note that the number $n$ of indices is usually referred to as the \emph{weight} of the HPL. Harmonic polylogarithms are a generalization of the classical polylogarithm functions, defined recursively by,
\beq\label{eq:lin_int}
\li{n}(z) = \int_0^z{\rd t\over t}\,\li{n-1}(t) {\rm~~and~~} \li{1}(z) = -\log(1-z)\,.
\eeq

Iterated integrals are well-known to form a shuffle algebra, and so in particular we can express a product of two harmonic polylogarithms of weight $n_1$ and $n_2$ as a linear combination of HPL's of weight $n_1+n_2$,
\beq\label{eq:shuffle}
H(\vec a_1;z) \, H(\vec a_2;z) \, = \, \sum_{\vec a=\vec a_1\uplus\vec a_2} H(\vec a;z) \, ,
\eeq
where $\vec a_1\uplus\vec a_2$ denotes the shuffle of the two weight vectors $\vec a_1$ and $\vec a_2$, 
\emph{i.e.}, all possible concatenations of $\vec a_1$ and $\vec a_2$ in which relative orderings of $\vec a_1$ 
and $\vec a_2$ are preserved.

Up to weight three, HPL's are known to be expressible through ordinary logarithms and the classical polylogarithms  $\li{n}$ only. Starting from weight four, not all HPL's can be expressed in terms of classical polylogarithms, and genuine new functions appear. For special classes of HPL's, however, it is possible to find closed expressions in terms of other functions. An example of this was already given in Eq.~\eqref{eq:H000}. Moreover, we have,
\beq\bsp
H(\vec s_n;z) = {(-s)^n\over n!}\log^n(1-s\,z) {\rm~~and~~} H(\vec0_{n-1},1;z) = \li{n}(z)\,,
\esp\eeq
\emph{i.e.}, harmonic polylogarithms contain the classical polylogarithms as special cases.

Apart from the shuffle relation~\eqref{eq:shuffle}, HPL's satisfy various intricate functional equations, relating HPL's with different arguments among each other. As an example, the functional equation relating HPL's with opposite arguments reads,
\beq
H(\vec a;-z) = (-1)^p \,H(-\vec a;z)\,,
\eeq
where $p$ denotes the number elements in $\vec a$ equal to $\pm1$. Furthermore, it is always possible to express harmonic polylogarithms of the form $H(\vec a;1/z)$ as a linear combination of HPL's of the form $H(\vec a;z)$. This allows in particular to analytically continue the harmonic polylogarithms outside the unit disc, a property that will be used later in the numerical implementation of the HPL's into the {\sc Chaplin} library. Additional relations among HPL's with related arguments have been presented in Ref.~\cite{Remiddi:1999ew, Maitre:2005uu}. 

Let us conclude this section by discussing some special values of the argument $z$ for which the HPL's can be expressed in terms of known transcendental numbers. First, it is easy to see that, unless $\vec a = \vec 0_n$, all HPL's vanish for $z=0$. Second, if the argument $z$ of a harmonic polylogarithm is $\pm1$, then it can be expressed in terms of so-called \emph{colored multiple zeta values}\footnote{Also known as \emph{alternating multiple zeta values} or \emph{Euler-Zagier sums.}} (CMZV's), 
\beq\bsp
H&(\underbrace{0,\ldots,0}_{m_1-1},s_1,\ldots,\underbrace{0,\ldots,0}_{m_k-1},s_k;1)\\
& = (-1)^{m+p}\,\zeta(m_1,\ldots,m_k;s_1,s_2/s_1,\ldots,s_k/s_{k-1})\,,
\esp\eeq
with $s_i=\pm1$ and $m=m_1+\ldots+m_k$, and $p$ denotes the number of elements in $\{s_i\}$ equal to +1.
The CMZV's are defined by nested sums,
\beq\label{eq:CMZV_def}
\zeta(m_1,\ldots,m_k;\sigma_1,\ldots,\sigma_k) = \sum_{0<n_1<n_2<\dots <n_k} \frac{\sigma_1^{n_1} \sigma_2^{n_2} \cdots \sigma_k^{n_k} }{n_1^{m_1} n_2^{m_2} \cdots n_k^{m_k} }\,,
\eeq
with $\sigma_i=\pm1$. CMZV's are convergent if and only if $(m_1,\sigma_1)\neq(1,1)$. It follows then immediately that HPL's of the form $H(\pm1,\vec a;\pm1)$ are in general divergent\footnote{In some cases the divergence can be tamed, \emph{e.g.}, $\lim_{z\to0}H(1,0;z)  = -\zeta_2$.}.


\section{Reduction to basis functions}
\label{reductionchapter}

From the previous section it is clear that many HPL's are not independent functions, but they are related among themselves by various intricate relations. In order to achieve an efficient numerical implementation, it is desirable to have as few independent functions as possible, \emph{i.e.}, we would like to resolve all the identities in order to arrive at a minimal set of functions, which are `as simple as possible' and through which all other HPL's can be expressed.

All the functional equations among harmonic polylogarithms (or, more generically, among multiple polylogarithms) can be resolved through the so-called symbol calculus. At the heart of the symbol calculus is the so-called \emph{symbol map}~\cite{Goncharov:2009}, a linear map that associates to an HPL of weight $n$ a tensor of rank $n$. As an example, the tensor associated to the classical polylogarithm $\textrm{Li}_n(z) = H(\vec0_{n-1},1;z)$ reads,
\beq
\begin{cal}S\end{cal}(\textrm{Li}_n(z)) = -(1-z)\otimes\underbrace{z\otimes\ldots\otimes z}_{(n-1) \textrm{ times}}\,.
\eeq
Furthermore, the symbol maps products that appear inside the tensor product to a sum of tensors,
\beq
\ldots\otimes(X\cdot Y)\otimes\ldots = 
\ldots\otimes X\otimes\ldots+
\ldots\otimes Y\otimes\ldots\,.
\eeq
It is conjectured that all the functional identities among (multiple) polylogarithms are mapped under the symbol map $\cS$ to algebraic relations among the tensors. Hence, the symbol calculus provides an effective way to resolve all the functional equations among (a certain class of) multiple polylogarithms.

In Ref.~\cite{Gangl:2011}, the symbol map was used to obtain a set of basis functions through which all HPL's up to weight four can be expressed. The basis functions obtained in Ref.~\cite{Gangl:2011} read,
\begin{itemize}
\item for weight one,
\beq\label{eq:B1}
\cB_1^{(1)}(z)= \ln z, \quad \cB_1^{(2)}(z)= \ln (1-z),\quad \cB_1^{(3)}(z)= \ln (1+z)\,,
\eeq
\item for weight two,
\beq\label{eq:B2}
\cB_2^{(1)}(z)= \textrm{Li}_2(z), \quad \cB_2^{(2)}(z)= \textrm{Li}_2(-z),\quad \cB_2^{(3)}(z)= \textrm{Li}_2\left({1-z\over 2}\right)\,,
\eeq
\item for weight three,
\beq\bsp\label{eq:B3}
&\cB_3^{(1)}(z)= \textrm{Li}_3(z), \quad \quad\quad\quad\cB_3^{(2)}(z)= \textrm{Li}_3(-z),\quad \quad\quad\cB_3^{(3)}(z)= \textrm{Li}_3(1-z),\\
& \cB_3^{(4)}(z)= \textrm{Li}_3\left({1\over1+z}\right)\,,\,\quad\cB_3^{(5)}(z)= \textrm{Li}_3\left({1+z\over2}\right), \quad  \cB_3^{(6)}(z)= \textrm{Li}_3\left({1-z\over2}\right),\\
&  \cB_3^{(7)}(z)= \textrm{Li}_3\left({1-z\over1+z}\right), \quad  \,\cB_3^{(8)}(z)= \textrm{Li}_3\left({2z\over z-1}\right)\,,
\esp\eeq
\item for weight four,
\beq\bsp\label{eq:B4}
&\cB_4^{(1)}(z)= \textrm{Li}_4(z), \quad\quad\quad\,\,\,\, \cB_4^{(2)}(z)= \textrm{Li}_4(-z),\\
& \cB_4^{(3)}(z)= \textrm{Li}_4(1-z), \quad\quad\cB_4^{(4)}(z)= \textrm{Li}_4\left({1\over1+z}\right),\\
&\cB_4^{(5)}(z)= \textrm{Li}_4\left({z\over z-1}\right), \quad  \cB_4^{(6)}(z)= \textrm{Li}_4\left({z\over z+1}\right),\\
&\cB_4^{(7)}(z)= \textrm{Li}_4\left({1+z\over2}\right), \quad  \cB_4^{(8)}(z)= \textrm{Li}_4\left({1-z\over2}\right),\\
&  \cB_4^{(9)}(z)= \textrm{Li}_4\left({1-z\over1+z}\right), \quad  \cB_4^{(10)}(z)= \textrm{Li}_4\left({z-1\over z+1}\right),\\
&\cB_4^{(11)}(z)= \textrm{Li}_4\left({2z\over z+1}\right),\quad \cB_4^{(12)}(z)= \textrm{Li}_4\left({2z\over z-1}\right),\\
&\cB_4^{(13)}(z)= \textrm{Li}_4\left(1-z^2\right),\quad \cB_4^{(14)}(z)= \textrm{Li}_4\left({z^2\over z^2-1}\right),\\
&\cB_4^{(15)}(z)= \textrm{Li}_4\left({4z\over (z+1)^2}\right)\,.
\esp\eeq
\end{itemize}
All harmonic polylogarithms up to weight three can be expressed through the basis functions in Eq.~(\ref{eq:B1} - \ref{eq:B4}). Starting from weight four, we need to extend the set of functions by adjoining three new elements to the basis,
\beq\bsp\label{eq:B11}
\cB_4^{(16)}(z) = \textrm{Li}_{2,2}(-1,z),\quad& \cB_4^{(17)}(z) = \textrm{Li}_{2,2}\left({1\over 2},{2z\over z+1}\right),\\ \cB_4^{(18)}(z) &= \textrm{Li}_{2,2}\left({1\over 2},{2z\over z-1}\right)\,,
\esp\eeq
where $\li{2,2}$ denotes a two-variable multiple polylogarithm that cannot be expressed through classical polylogarithms only,
\beq
\li{2,2}(z_1,z_2) = \sum_{n_1=1}^\infty\sum_{n_2=1}^{n_1-1}{z_1^{n_1}\over n_1^2}\,{z_2^{n_2}\over n_2^2}\,.
\eeq
For practical purposes we find it more convenient to use a different set of multiple polylogarithms as basis functions than the one used in Ref.~\cite{Gangl:2011}. More specifically, we find it more convenient to perform a change of basis and replace the functions $\cB_4^{(i)}(z)$, for $i = 16, 17, 18$, by the functions $\tilde\cB_4^{(i)}(z)$, which are directly expressed as HPL's,
\beq\bsp\label{eq:B4p}
\tilde\cB_4^{(16)}(z) &\,= H(0,1,0,-1;z) = \cB_4^{(16)}(-z)\,,\\
\tilde\cB_4^{(17)}(z) &\,= H(0,1,1,-1;z)\,,\\
\tilde\cB_4^{(18)}(z) &\,= H(0,1,1,-1;-z)\,.
\esp\eeq
The set of the 32 functions $\cB_i^{(j)}(z)$ defines a basis through which all HPL's up to weight four can be expressed. As a consequence, any numerical code to evaluate this set of basis functions will automatically be able to evaluate all 120 HPL's up to weight four. Furthermore, as the basis functions only involve the two genuine multiple polylogarithms $H(0,1,0,-1;z)$ and $H(0,1,1,-1;z)$ besides the ordinary logarithms and the classical polylogarithms $\li{2}(z)$, $\li{3}(z)$ and $\li{4}(z)$, it is enough to have numerical routines for these latter functions. In this way we can reduce the problem of evaluating the 120 HPL's up to weight four to only a handful of non-trivial numerical routines. In the {\sc Chaplin} library, these routines consist of series expansions for the aforementioned functions that will be described in the next section.

Let us conclude this section by reviewing some of the properties of the basis functions $\cB_i^{(j)}(z)$ derived in Ref.~\cite{Gangl:2011}. First, it is easy to check that all the basis functions are real for $z\in[0,1]$. However, we stress that the expressions in Eq.~(\ref{eq:B1} - \ref{eq:B4p}) are strictly valid only for $z\in[0,1]$. While most of the expressions are valid everywhere throughout the unit disc, the analytic form of $\cB_4^{(13)}(z)$ valid on the whole interior of the disc reads~\cite{Gangl:2011},
  \beq\bsp\label{eq:Li41mz2}
&  \cB_4^{(13)}(z) \\
  &= \left\{\begin{array}{ll}
  \textrm{Li}_4(1-z^2)\,,  \quad\quad\quad\,\,\,\,\,\textrm{if Re}(z) >0 &\textrm{ or (\textrm{Re}}(z)=0 \textrm{ and Im}(z)\ge0)\,,\\ 
  \textrm{Li}_4(1-z^2)-{i\pi\over3}\,\sigma(z)\,\ln^3(1-z^2)\,,&\textrm{otherwise}\,,
  \end{array}\right.
  \esp\eeq
  where $\sigma(z) = \textrm{sign}(\textrm{Im}(z))$. 
 Second, since it is our goal to build a numerical code to evaluate HPL's for arbitrary complex arguments, we need  to analytically continue the basis functions outside the unit disc. In Ref.~\cite{Gangl:2011} inversion relations of the form
\beq
\cB_j^{(i)}(z) = \sum_{k,l}\,c_{ijkl}\,\cB_k^{(l)}\left({1\over z}\right)+\textrm{products of lower weight.}
\eeq
were derived that can be used for this purpose. Finally, we note that there is a subtlety in the basis function $\cB_4^{(15)}\left(z\right)$ when going from the interior to the exterior of the unit disc because $\cB_4^{(15)}(z)$ has a branch cut along the unit circle in the complex $z$-plane. In Ref.~\cite{Gangl:2011} it was shown that if we want $\cB_4^{(15)}(z)$ to be continuous and real for $z\in[0,1]$, we need to choose the following prescription for $|z|=1$,
\beq
\cB_4^{(15)}\left(z\right) = \textrm{Li}_4\left({4z\over(1+z)^2}+i\sigma(z)\varepsilon\right)\,.
\eeq

%

\section{Numerical evaluation of the basis functions}
\label{sec:series}
\subsection{Notations and conventions}
In the previous section we introduced a set of basis functions through which every HPL up to weight four can be expressed. The basis has the property that it only involves two types of new functions, besides the ordinary logarithm and the classical polylogarithms. These two new functions can be chosen to correspond to the two harmonic polylogarithms $H(0,1,0,-1;z)$ and $H(0,1,1,-1;z)$. 
In this section we present series expansions of these functions which are used inside {\sc Chaplin} to evaluate the basis functions.

Let us start by introducing some notations and conventions. As we will deal with series expansions, we define some operations on the coefficients of the series, \emph{i.e.}, on sequences of complex numbers.
For two sequences of complex numbers $a_n$ and $b_n$, $n\in \mathbb{N}$, we define their convolution product as the sequence $(a\ast b)_n$ defined by
\beq\label{eq:convolution_def}
(a\ast b)_n = \sum_{k=0}^n\binom{n}{k}\,a_k\,b_{n-k}\,.
\eeq
It is easy to check by manipulating the sum that this operation is associative, commutative and has the sequence $\varepsilon_n = \delta_{0,n}$, where $\delta_{i,j}$ is the Kronecker symbol, as a neutral element,
\beq\label{eq:convol_props}
a\ast (b\ast c) = (a\ast b)\ast c\,, \quad a\ast b = b\ast a\,, \quad a\ast \varepsilon = \varepsilon \ast a = a\,.
\eeq
The fact that $\varepsilon_n$ is a neutral element is obvious, and the commutativity follows immediately from changing the summation variable from $k$ to $n-k$. Associativity is less obvious, and is proved in Appendix~\ref{app:Associativity}.
Furthermore, this operation is compatible with the usual termwise addition and scalar multiplication  of sequences,
\beq
a\ast(b+c) = a\ast b + a\ast c\,,\quad (\kappa\cdot a)\ast b = \kappa\cdot (a\ast b)\,,
\eeq
where $a$, $b$ and $c$ are sequences of complex numbers and $\kappa$ is a constant complex number. The convolution product allows us to write the coefficients that appear in the product of two power series as the convolution of the coefficients of the individual factors, \emph{e.g.},
\beq
\left(\sum_{m=0}^\infty {a_m\over m!}\,x^m\right) \,\left(\sum_{n=0}^\infty {b_n\over n!}\,x^n\right)
=
\sum_{N=0}^\infty {(a\ast b)_N\over N!}\,x^N\,.
\eeq
Finally, for later convenience, we define for a given sequence $a_n$ of complex numbers the three new sequences $\mathring a_n$, $\bar a_n$ and $s(a)_n$ by
\beq\label{eq:bar_tilde}
\mathring a_n = {a_n\over n+1}\,,\quad \bar a_n = (-1)^n\,{a_n}\,,\quad s(a)_n=\left\{\begin{array}{ll}a_{n-1}\,, &\textrm{if } n\ge1\\ 0\,,&\textrm{otherwise}\end{array}\right.\,.
\eeq
The bar-operation allows us to define the coefficients of the series expansion of $f(-z)$ in terms of the coefficients of the series expansion of $f(z)$. More precisely, the two series expansions are related by
\beq
f(z) = \sum_{n=0}^\infty f_n\,z^n {\rm~~and~~} f(-z) = \sum_{n=0}^\infty \bar{f}_n\,z^n\,.
\eeq
We also define the composition of the `$\circ$'-operation and the convolution and shift operations,
\beq
(a\circledast b)_n = (a\ast b)^\circ_n = {(a\ast b)_n\over n+1} {\rm~~and~~}\mathring{s}(a)_n = {s(a)_n\over n+1}\,.
\eeq
Note that the $\circledast$-operation is commutative, but not associative. For later convenience we introduce the following convention,
\beq
a\circledast b\circledast c \equiv a\circledast(b\circledast c)\,.
\eeq
The operations on sequences of complex numbers we just defined allow us to write the coefficients appearing in the series expansion of the basis functions in a compact closed form. The sequences of complex numbers that appear inside these closed-form expressions are well-known sequences of (rational) numbers which we recall in the following.
 
\begin{enumerate}
\item {\bf (Shifted) $\zeta$ values}: 
\beq
\zeta^{(k)}_n = \left\{\begin{array}{ll}
\zeta_{k-n}, &\textrm{if } k-n \neq1\,,\\
H_{k-1}, & \textrm{if } k-n = 1\,,
\end{array}\right.
\eeq
where $\zeta_m\equiv\zeta(m)$ denotes the Riemann $\zeta$ function and $H_m$ the $m$-th harmonic number,
\beq
\zeta(z) = \sum_{n=1}^\infty{1\over n^z} {\rm~~and~~} H_m = \sum_{n=1}^m{1\over n}\,.
\eeq
Note that $\zeta_n^{(k)}$ is rational for $n\ge k$ and transcendental of weight $k-n$ otherwise.
\item {\bf Bernoulli numbers:} The Bernoulli numbers $B_n$ are defined through the generating series
\beq\label{eq:Bernoulli}
{z\over e^z-1} = \sum_{n=0}^\infty B_n\,{z^n\over n!}\,.
\eeq
Note that $B_{2n+1}=0$, $\forall n\in\mathbb{N}$. The Bernoulli numbers are related to $\zeta$ values via the formula,
\beq\label{eq:Bernoulli_zeta}
\zeta_0 = B_1=-{1\over2} {\rm~~and~~} \zeta_{-n} = -{B_{n+1}\over n+1} \textrm{ for } n\ge 1\,.
\eeq
\item {\bf Genocchi numbers:} The Genocchi numbers are defined through the generating series
\beq\label{eq:Gen_Func_Genocchi}
{2t\over 1+e^t} = \sum_{n=0}^\infty{G_n\over n!}\,t^n\,.
\eeq
They are related to the Bernoulli numbers via
\beq
G_n = 2\,(1-2^n)\,B_n\,.
\eeq
Note that $G_0 = G_{2n+1} = 0$.
\vskip 0.3cm
\item {\bf Polylogarithms in half-integer values:} The series expansion of the basis functions also involve the following sequences of numbers
\beq
\ell_n = (-1)^n\,\textrm{Li}_{-n}\left({1\over2}\right)\,.
\eeq
The sequence $\ell_n$ admits the generating function,
\beq
{1\over 2e^z-1} = \sum_{n=0}^\infty\ell_n\,{z^n\over n!}\,.
\eeq
\end{enumerate}

\subsection{Numerical evaluation of classical polylogarithms}
In this section we discuss series expansions of classical polylogarithms
that can be used to obtain reliable numerical results for these functions (at least in some regions of the complex plane).
In particular, truncated versions of these series are used by {\sc Chaplin} to evaluate the classical polylogarithms.
As the goal of {\sc Chaplin} is to provide an efficient way to evaluate harmonic polylogarithms for arbitrary complex arguments, we divide the problem into two regions: for complex numbers $z$ with $|z|\le1$ we use the series expansions to evaluate the basis functions numerically, whereas points with $|z|>1$ are mapped back into the interior of the unit disc using the inversion formul\ae\ for the classical polylogarithms. Hence, from here on we will only concentrate on complex number $z$ with $|z|\le1$.

Classical polylogarithms can be expanded into a power series around $z=0$,
\beq\label{eq:Lin_power_series}
\li{n}(z) = \sum_{k=1}^\infty{z^k\over k^n}\,.
\eeq
Even though this series is convergent for $|z|<1$, the convergence is rather slow. A faster convergence can be achieved by using the so-called Bernoulli substitution~\cite{Vollinga:2004sn}, which consists in expanding $\li{n}(z)$ into a series in \mbox{$\log(1-z)$}. 
While this expansion converges much faster for $|z|\ll1$ than the Taylor expansion~\eqref{eq:Lin_power_series}, it fails to produce reliable results when $z$ approaches 1. In Ref.~\cite{CLZ} an alternative expansion of the classical polylogarithms into a series in $\log z$ was derived. In this case the convergence is fast inside an annulus around $z=0$, but fails to converge for $|z|\ll1$. The strategy seems thus clear: we can split the interior of the unit disc into two distinct regions, and in each region one of the two series expansions converges quickly. Similar expansions can also be derived for the two remaining basis functions, $H(0,1,0,-1;z)$ and $H(0,1,1,-1;z)$ (and in principle for every HPL) and are discussed in the rest of this section. We start by deriving in detail the expansions of the dilogarithm, because even though these results are well-known, the techniques used in the derivation will be the starting point for the higher-weight cases.

\paragraph*{Series expansions of $\li{2}(z)$.}
Let us start by deriving the expansion of $\li{2}(z)$ into a series in $\log(1-z)$. Letting $x=-\log(1-z)$, this is equivalent to finding the Taylor series expansion of the function $\li{2}(1-e^{-x})$. We start from the integral representation~\eqref{eq:lin_int} and we get,
\beq
\li{2}(1-e^{-x}) = \int_0^{1-e^{-x}}{\rd t\over t}\li{1}(t)=- \int_0^{1-e^{-x}}{\rd t\over t}\log(1-t)\,.
\eeq
Performing the change of variables $t=1-e^{-t'}$ and using Eq.~\eqref{eq:Bernoulli}, we obtain,
\beq
\li{2}(1-e^{-x}) = -\int_0^{x}{e^{-t'}\rd t'\over 1-e^{-t'}}\,(-t') = \int_0^x\rd t'\,{t'\over e^{t'}-1} = \sum_{k=0}^\infty{B_k\over(k+1)!}\,x^{k+1}\,,
\eeq
or equivalently in terms of the original variable $z$,
\beq\label{eq:Bernoulli_Li2}
\li{2}(z) = \sum_{k=0}^\infty{B_k\over (k+1)!}\,(-\log(1-z))^{k+1}\,.
\eeq
The series expansion~\eqref{eq:Bernoulli_Li2} converges rather quickly inside a disc around $z=0$ of radius $R<1$ (the precise value of $R$ used by {\sc Chaplin} will be given in the next section). In the remaining annulus $R<|z|<1$ the dilogarithm admits a series expansion in $\log z$~\cite{CLZ},
\beq\label{eq:Li2_CLZ}
\li{2}(z) = -\log z\,\log(-\log z) + \sum_{k=0}^\infty{\zeta_k^{(2)}\over k!}\,\log^kz\,.
\eeq
Let us sketch the derivation of Eq.~\eqref{eq:Li2_CLZ}. Letting $x=\log z$, we start from the integral representation of the dilogarithm and perform the change of variable $t = e^{t'}$,
\beq\label{eq:Li2_CLZ_derivation}
\li{2}(e^{x}) = \zeta_2+\int_1^{e^{x}}{\rd t\over t}\,\li{1}(t) = \zeta_2 +\int_0^x\rd t'\,\li{1}(e^{t'})\,.
\eeq
In order to proceed, we need the Taylor expansion of $\li{1}(e^{x}) = -\log(1-e^{x})$. Using the integral representation of $\li{1}$ as well as Eq.~\eqref{eq:Bernoulli}, we obtain,
\beq\bsp\label{eq:Li1_CLZ_derivation}
\li{1}(e^{x}) &\,= \int_0^{e^{x}}{\rd t\over 1-t} = \lim_{\varepsilon\to0}\left[-\log(1-e^{\varepsilon}) -\int_\varepsilon^x{\rd t'\over t'}\,{(-t')\over e^{-t'}-1}\right]\\
&\,= \lim_{\varepsilon\to0}\left[-\log(1-e^{\varepsilon}) - \int_\varepsilon^x{\rd t'\over t'} -\sum_{n=1}^\infty{B_n\over n!} \int_\varepsilon^x\rd t'\,(-t')^{n-1}\right]\,.
\esp\eeq
The last term in Eq.~\eqref{eq:Li1_CLZ_derivation} is finite, whereas the logarithmic divergences cancel between the first two terms,
\beq\bsp
\lim_{\varepsilon\to0}&\left[-\log(1-e^{\varepsilon}) - \int_\varepsilon^x{\rd t'\over t'}\right] 
=
\lim_{\varepsilon\to0}\left[-\log(-\varepsilon +\begin{cal}O\end{cal}(\varepsilon^2)) - \log (-x)+\log(-\varepsilon)\right] \\
&\,=
\lim_{\varepsilon\to0}\left[-\log(1+\begin{cal}O\end{cal}(\varepsilon)) - \log(-x)\right] 
=-\log(-x)\,.
\esp\eeq
Hence, using Eq.~\eqref{eq:Bernoulli_zeta} and the fact that $\zeta_{-n} = 0$ for even $n$, we get,
\beq\bsp\label{eq:Li1_CLZ_derivation_2}
\li{1}(e^{x}) &\,= -\log(-x) -\sum_{n=1}^\infty{B_n\over n!} {(-x)^{n}\over n} = -\log(-x) +\zeta_0\,x+\sum_{n=1}^\infty{\zeta_{-n}\over (n+1)!}\, (-x)^{n+1}\\
&\,= -\log(-x) +\sum_{n=0}^\infty{\zeta_{-n}\over (n+1)!}\, x^{n+1}\,.
\esp\eeq
Inserting Eq.~\eqref{eq:Li1_CLZ_derivation_2} into Eq.~\eqref{eq:Li2_CLZ_derivation} and integrating term by term immediately reproduces Eq.~\eqref{eq:Li2_CLZ} (after identification $x=\log z$).
Note the appearance of the nested logarithm, $\log(-\log z)$ in Eq.~\eqref{eq:Li2_CLZ}, which seems to be divergent for $z$ close to 1. It is however easy to check that
\beq
\lim_{z\to1^-}\log z\,\log(-\log z) = 0\,,
\eeq
and so the whole expression is well behaved.

\paragraph*{Series expansions of $\li{n}(z)$, $n>2$}.
The derivations of the series expansions of $\li{2}(z)$ presented in the previous section are by no means restricted to the weight two case, but we can repeat exactly the same steps for classical polylogarithms of arbitrary weight.
In Ref.~\cite{CLZ} the following more general version of Eq.~\eqref{eq:Li2_CLZ} was proven,
\beq\label{eq:Lin_CLZ}
\li{n}(z) = -{1\over (n-1)!}\,\log^{n-1}z\,\log(-\log z) + \sum_{k=0}^\infty{\zeta^{(n)}_k\over k!}\,\log^kz\,.
\eeq
The proof goes by recursion in the weight, and we refer to Ref.~\cite{CLZ} for more details.

The generalization of Eq.~\eqref{eq:Bernoulli_Li2} to arbitrary weight is simply given by,
\beq\label{eq:Lin_Bernoulli}
\li{n}(z) = \sum_{k=0}^\infty{1\over (k+1)!}\,[B\ast(\underbrace{B\circledast\ldots\circledast B\circledast\mathring{B}}_{n-2 \textrm{ times}})]_k\,(-\log(1-z))^{k+1}\,.
\eeq
The proof goes by recursion in the weight. We have shown in the previous section that Eq.~\eqref{eq:Lin_Bernoulli} is true for $n=2$. If we assume in addition Eq.~\eqref{eq:Lin_Bernoulli} true up to weight $n$, then we obtain,
\beq\bsp
\li{n+1}(1-e^{-x}) &\,= \int_0^{1-e^{-x}}{\rd t\over t}\li{n}(t) \\
&\,= \sum_{l=0}^\infty{1\over (l+1)!}\,[B\ast(\underbrace{B\circledast\ldots\circledast B\circledast\mathring{B}}_{n-2 \textrm{ times}})]_l\int_0^x{t'\,\rd t'\over e^{t'-1}}\,t'^l\\
&\,= \sum_{l=0}^\infty{1\over l!}\,(\underbrace{B\circledast\ldots\circledast B\circledast\mathring{B}}_{n-1 \textrm{ times}})_l\sum_{m=0}^\infty{B_m\over m!}\,{x^{l+m+1}\over l+m+1}\\
&\,=\sum_{k=0}^\infty{1\over k!}\,[B\ast(\underbrace{B\circledast\ldots\circledast B\circledast\mathring{B}}_{n-1 \textrm{ times}})]_k\,{x^{k+1}\over k+1}\,.
\esp\eeq
The series expansions in $\log z$ and $\log(1-z)$ for the classical polylogarithms are enough to evaluate all basis functions, except $\tilde B_4^{(i)}$, for $i\in\{16,17,18\}$, numerically in a fast and reliable way. For $\tilde B_4^{(i)}$, for $i\in\{16,17,18\}$ we need to extend these series expansions beyond the case of classical polylogarithms.

\subsection{Series expansions of $H(0,1,0,-1;z)$ and $H(0,1,1,-1;z)$}
In this section we present the analogues of the series expansions in $\log(1-z)$ and $\log z$ of the previous section for the three remaining basis functions that cannot be expressed in terms of classical polylogarithms only, namely $\tilde B_4^{(i)}$, for $i\in\{16,17,18\}$. As the derivation of the expansions follows exactly the same lines as for the classical polylogarithms, we content ourselves to present the results and refer to Appendix~\ref{app:proofs} for details.

Let us start by deriving the expansion of $\tilde B_4^{(i)}$, $i\in\{16,17,18\}$ into a series in $\log(1-z)$. We can obtain these expansions as a corollary of a more general result: If a function $\cH(z)$ admits a Taylor expansion of the form
\beq
\cH(1-e^{-x}) = \sum_{n=0}^\infty{h_n\over (n+1)!}\,x^{n+1}\,,
\eeq
then for $a\in\{-1,0,1\}$ the functions $\cH_a(z)$ defined by
\beq\label{eq:cH_int}
\cH_a(z) = \int_0^z\rd t\,f(a;z)\,\cH(t)\,,
\eeq
where $f(a;z)$ was defined in Eq.~\eqref{eq:faz},
admit the following Taylor expansions
\beq\bsp\label{eq:HPL_Log(1-x)_recursive}
\cH_1(1-e^{-x}) &\,= \sum_{n=0}^\infty{s(h)_n\over (n+1)!}\,x^{n+1}=\sum_{n=0}^\infty{h_n\over (n+2)!}\,x^{n+2}\,,\\
\cH_0(1-e^{-x}) &\,= \sum_{n=0}^\infty{(B\ast\mathring{h})_n\over (n+1)!}\,x^{n+1}\,,\\
\cH_{-1}(1-e^{-x}) &\,= \sum_{n=0}^\infty{(\ell\ast s(h))_n\over (n+1)!}\,x^{n+1}\,.
\esp\eeq
To prove these identities we start from the integral representation~\eqref{eq:cH_int} and, after having performed the change of variable $t=1-e^{-t'}$, we insert the series expansion for the integrand and integrate term by term, which yields immediately Eq.~\eqref{eq:HPL_Log(1-x)_recursive}.
Combined with the expansions for HPL's of weight one, 
Eq.~\eqref{eq:HPL_Log(1-x)_recursive} allows us in principle to recursively expand HPL's of arbitrary weight into a series in $\log(1-z)$. A similar iterative procedure was already described in Ref.~\cite{Vollinga:2004sn}. As the classical polylogarithms are just a special case of the harmonic polylogarithms, Eq.~\eqref{eq:Lin_Bernoulli} can be seen as a special solution of Eq.~\eqref{eq:HPL_Log(1-x)_recursive} for $H(\vec0_{n-1},1;1-e^{-x})$. 
Moreover, we can easily read off from Eq.~\eqref{eq:HPL_Log(1-x)_recursive} the corresponding expansions of $H(0,1,0,-1;z)$ and $H(0,1,1,-1;z)$,
\beq\bsp\label{eq:Logx_H}
H(0,1,0,-1;z) &\,= \sum_{n=0}^\infty{[B\ast\mathring{s}(B\ast\mathring{\ell})]_n\over(n+1)!}\,(-\log(1-z))^{n+1}\,,\\
H(0,1,1,-1;z) &\,= \sum_{n=0}^\infty{[B\ast\mathring{s}^2(\ell)]_n\over(n+1)!}\,(-\log(1-z))^{n+1}\,,\\
\esp\eeq
with $\mathring{s}^2(\ell)_n=\ell_{n-2}/(n+1)$\,. The expansions~\eqref{eq:Logx_H} are used inside {\sc Chaplin} to evaluate the basis functions $\tilde B_4^{(i)}(z)$, $i\in\{16,17,18\}$ for $z$ close to the origin. Similar to the case of the classical polylogarithms, these series converge rather slowly if $z$ is close to 1. We therefore need additional expansions in $\log z$ that have a good convergence behavior in that region. 

 In an annulus inside the unit disc, the basis functions $\tilde B_4^{(i)}(z)$, $i\in\{16,17,18\}$ can be expanded into a series in $\log z$. As these basis functions are entirely expressed through $H(0,1,0,-1;z)$ and $H(0,1,1,-1;z)$, it is enough to find the expansions for these cases. We obtain,
\beq\bsp\label{eq:HPL_Logx}
H&(0,1,0,-1;z) =-4\, \text{Li}_4\left(\frac{1}{2}\right)-\frac{5}{2}\, \zeta_3\, \log2+\frac{17 \pi ^4}{480}-\frac{1}{6}\,\log ^42\\
&+\frac{\pi ^2 }{6}\, \log ^22-{5\over8}\,\zeta_3\,\log z+{\pi^2\over 6}\,\log2\,\log z
+{\pi^2\over 12}\,\li{2}(z)\\
&+\log2\,\log z\,\li{2}(z)-2\log2\,\li{3}(z)
+{1\over 2}\sum_{n=0}^\infty{(\overline{B}\ast\mathring{\gamma})_n\over(n+2)!}\,\log^{n+2}z\,,\\
H&(0,1,1,-1;z) = \text{Li}_4\left(\frac{1}{2}\right)-\frac{1}{8}\, \zeta_3\, \log 2+\frac{\pi ^4}{720}+\frac{1}{24}\,\log ^42\\
&+\left(\frac{1}{2}\, \log ^22-\frac{\pi ^2 }{12}\right)\,\text{Li}_2(z)+\left(\frac{7}{8}\, \zeta_3+\frac{1}{6}\, \log ^32-\frac{\pi^2}{12}\, \log 2\right)\, \log z\\
&-{1\over 2}\,\sum_{n=0}^\infty{(\overline B\ast\beta)_n\over(n+1)!}\,{1\over n}\,\log^{n+1}z\,,
\esp\eeq
with
\beq
\gamma_n={\overline G_n\over n}{\rm~~and~~}\beta_n={(\overline B\ast\gamma)_n\over n}\,.
\eeq
The proof of the Eq.~\eqref{eq:HPL_Logx} is sketched in Appendix~\ref{app:proofs}. A similar expansion in the particular case of $H(1,\vec 0_n,1;z)$ has already been considered in Ref.~\cite{Gangl:Num}. We stress that the relations~\eqref{eq:HPL_Logx} are only valid in the region $\textrm{Re}(z)\ge0$ and $R\le|z|\le1$, for some $R$ (we give a precise value for $R$ in the next section when discussing the implementation of these expansions into {\sc Chaplin}). For $\textrm{Re}(z)<0$, we proceed in following way,
\begin{itemize}
\item for $H(0,1,0,-1;z)$, we map the problem to $\textrm{Re}(z)>0$ via the functional equation
\beq\bsp
H&(0,1,0,-1;z) = -H(0,-1,0,1;-z)\\
&=
H(0,1,0,-1;-z)-\frac{1}{2}\,\text{Li}_4\left(1-z^2\right)+
\frac{1}{2}\, \text{Li}_4\left(\frac{z^2}{z^2-1}\right)\\
&-2\, \text{Li}_4\left(\frac{1}{1-z}\right)+2\, \text{Li}_4(-z)+2\, \text{Li}_4(z)-2\, \text{Li}_4\left(\frac{z}{z+1}
\right)\\
&-2\, \text{Li}_4\left(\frac{z}{z-1}\right)+2\, \text{Li}_4(1+z)-2\, \text{Li}_3(-z)\, \log (1-z)\\
&-2\, \text{Li}_3(z)\, \log (1+z)-\frac{3}{2} \,
\zeta_3\, \log (1-z)-\frac{3}{2}\, \zeta_3\, \log (1+z)\\
&-\frac{7}{48} \,
\log ^4(1-z)-\frac{1}{16} \log ^4(1+z)+\frac{1}{6}\, \log (-z)\, \log 
^3(1-z)\\
&+\frac{1}{12}\, \log (1+z)\, \log^3(1-z)+\frac{1}{12}\, \log 
^3(1+z)\, \log (1-z)
\esp\eeq
\beq\bsp
\phantom{H}&+\frac{1}{6}\, \log (-z)\, \log ^3(1+z)+\frac{1}{8} \,
\log ^2(1+z)\, \log ^2(1-z)\\
&-\frac{1}{2}\, \log (-z)\, \log (1+z)\, \log 
^2(1-z)+\frac{5\pi ^2 }{24}\, \log ^2(1-z)\\
&-\frac{1}{2}\, \log (-z)\, \log 
^2(1+z)\, \log (1-z)-\frac{\pi^2}{8} \, \log ^2(1+z)\\
&-\text{Li}_2(-z) 
\,\text{Li}_2(z)+\frac{\pi^2}{12} \, \log (1+z)\, \log (1-z)+\frac{\pi ^4}{180} \,.
\esp\eeq
\item for $H(0,1,1,-1;z)$, we use the following expansion, valid for $\textrm{Re}(z)<0$,
\beq\bsp
H&(0,1,1,-1;z) = 
2\, \text{Li}_4\left(\frac{1}{2}\right)-\frac{7 \pi ^4}{360}+\frac{1}{12}\,\log ^42+\frac{\pi^2}{12} \, \log ^22\\
&+\left(\frac{1}{8}\,\zeta_3-\frac{1}{6}\,\log ^32+\log2\right)\,\log z-\left(\frac{\pi ^2}{12}-\frac{1}{2}\,\log ^22\right)\,\li{2}(-z)\\
&-{1\over4}\,\sum_{n=0}^\infty{(\overline G\ast \bar{g}^{(1)})_n\over(n+1)!}\,{1\over n}\,\log^{n+1}(-z)\,\left(\log(-\log (-z))-{1\over n+1}-{1\over n}\right)\\
&+{1\over4}\,\sum_{n=0}^\infty{(\overline G\ast \bar{g}^{(2)})_n\over(n+1)!}\,{1\over n}\,\log^{n+1}(-z)+{1\over4}\sum_{n=0}^\infty{[\overline{G}\ast(\overline{G}\circledast\xi)]_n\over(n+2)!}\,\log^{n+2}(-z)\,,
\esp\eeq
with
\beq
g^{(i)}_n={G_n\over n^i} {\rm~~and~~} \xi_n = \mathring{\zeta}_{-n}\,.
\eeq
\end{itemize}
At this stage we have all the ingredients to evaluate all harmonic polylogarithms up to weight four numerically for arbitrary complex arguments. To this effect, we have implemented the decomposition of HPL's to the basis defined by Eq.~(\ref{eq:B1} - \ref{eq:B4p}) and the series expansions presented in this section into the {\sc Chaplin} library, which we will present in the next section.

%
\section{The Fortran library {\sc Chaplin}}
\label{sec:chaplin}
\subsection{Installation}
The {\sc Chaplin} code is available as a \verb+.tar+ archive on the website\\
 \url{http://projects.hepforge.org/chaplin/} .

Once unpacked, the code can be compiled via
\begin{verbatim}
 ./configure
 make install
\end{verbatim}
As a result, both a static and a shared library are created in the directory \verb+/usr/lib+, which may require root privileges. The directory which the library is installed in can be changed during configuration via
\begin{verbatim}
./configure --prefix=/"path to Chaplin/"
\end{verbatim}
This feature allows to create the library even without root privileges.

We advocate the use of the shared library when linking {\sc Chaplin} to a {\tt Fortran} code, such that the routines needed by the program will only be called during run-time. Static linking, on the contrary, puts all the {\sc Chaplin} functions into the executable {\sc Chaplin} is linked against, which might result in long compilation times and rather large executables.
\subsection{Running {\sc Chaplin}}
Once compiled, the {\sc Chaplin} library can be linked to other programs in the same way as any other library.
This enables the user to call the numerical routines of {\sc Chaplin} from within his/her own program.
The function calls to the numerical routines of {\sc Chaplin} are done via the following functions,
\begin{verbatim}
 double complex HPL1(n1, z)
 double complex HPL2(n1, n2, z)
 double complex HPL3(n1, n2, n3, z)
 double complex HPL4(n1, n2, n3, n4, z)
\end{verbatim}
with \verb+ni+ $\in \{-1,0,1\}$ and \verb+z+ being any \verb+double complex+ number. The return value is $H(\vec n; z)$, \emph{i.e.}, the HPL with weight vector $\vec n$ at the point $z$. Alternatively, in case only the real and/or imaginary parts of an HPL are needed, the user might find the following functions convenient,
\begin{verbatim}
 double precision HPL2real(n1, n2, xr, xi)
 double precision HPL2im(n1, n2, xr, xi)
\end{verbatim}
and similarly for \verb+HPL3, HPL4+. The variables \verb+xr, xi+ are \verb+double precision+ variables, denoting the real and imaginary parts of the argument of the HPL. Note that these functions are useful to call {\sc Chaplin} from within a \verb|C++| program, where complex numbers are not natively supported.

\begin{figure}[!t]
 \centering
  \includegraphics[width=0.8\textwidth]{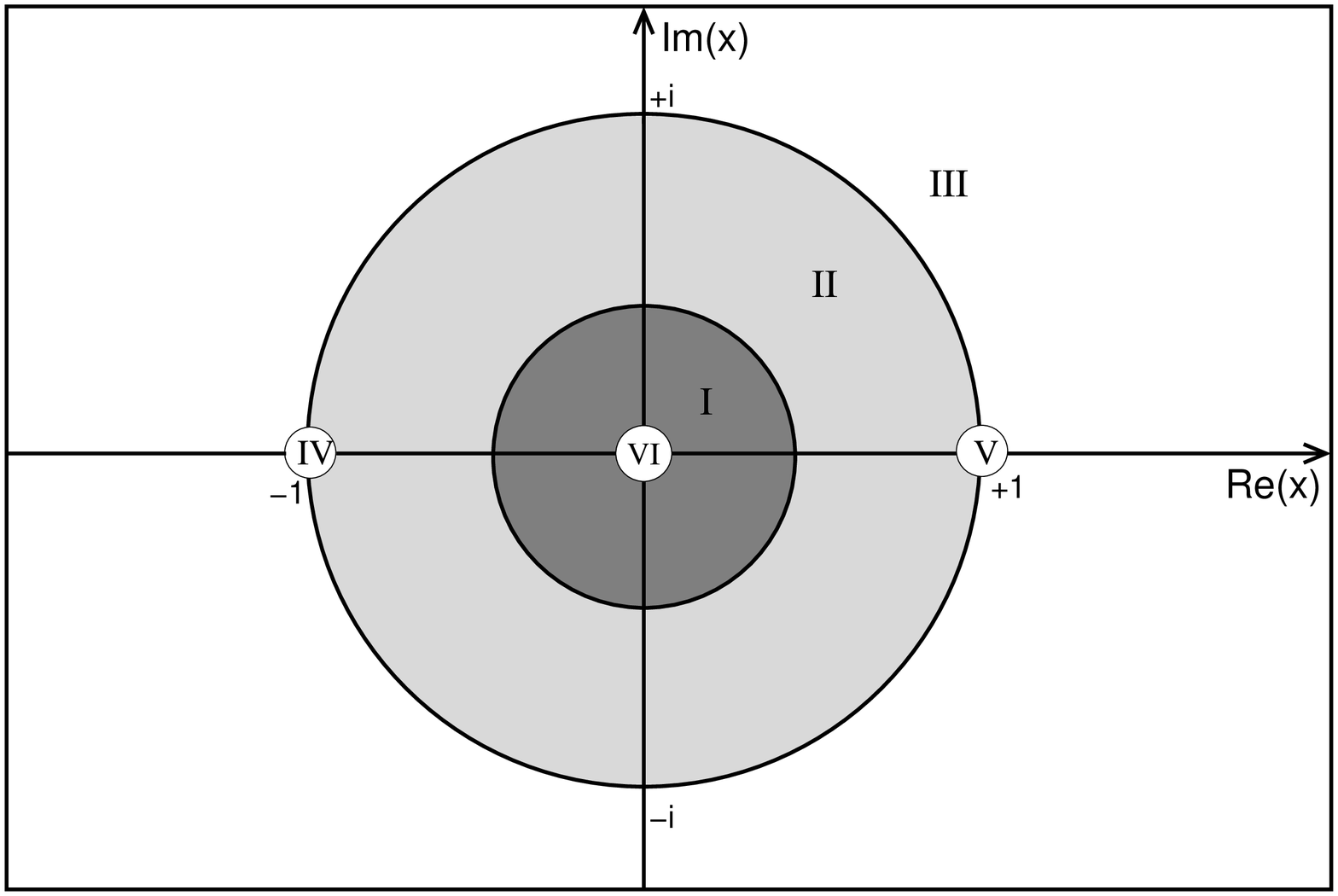}
  \caption{The different regions of the complex plane used inside the {\sc Chaplin} library.}  
  \label{fig}
\end{figure}

When one of the aforementioned functions is called, {\sc Chaplin} starts by decomposing the corresponding HPL internally into the basis of 32 functions described in Section~\ref{reductionchapter}. In a second step, {\sc Chaplin} proceeds to the numerical evaluation of the individual basis functions appearing in the decomposition.
The numerical routines called to this effect are different depending on the value of the argument $z$ of the HPL. More precisely, the complex plane is divided into six regions (See Fig.~\ref{fig}),
\begin{itemize}
\item Region I: inside an annulus $0.025<|z|\le 0.3$, the basis functions are evaluated by using the expansions in $\log(1-z)$ presented in Section~\ref{sec:series}.
\item Region II: inside an annulus $0.3\le|z|\le1$, the basis functions are evaluated by using the expansions in $\log z$ presented in Section~\ref{sec:series}.
\item Region III: points outside the unit disc, $|z|>1$, are mapped back to the interior of the unit disc via inversion relations.
\item Regions IV \& V: The basis of Section~\ref{reductionchapter} involves functions that are logarithmically divergent at $\pm1$, leading to spurious singularities in the basis expansion. To avoid numerical instabilities caused by these spurious singularities, we use Taylor expansions close to $z=\pm1$ to evaluate the individual HPL's without proceeding to a decomposition into basis functions.
\item Region VI: In order to achieve a good numerical precision close to the origin of the complex plane, we use Taylor expansions in a disc $|z| < 0.025$ without proceeding to a decomposition into basis functions.
\end{itemize}

At the end of this procedure, {\sc Chaplin} returns the numerical value of the HPL given as an input. In case the numerical evaluation of a divergent quantity is attempted (\emph{e.g.}, $H(\vec 0_n; 0)$ or $H(\pm1,\vec a;\pm1)$) an exception is thrown and the evaluation is aborted. Note that for real values of the argument, {\sc Chaplin} uses the `$+i\varepsilon$' prescription, \emph{i.e.}, for $z\in\mathbb{R}$, $H(\vec a;z)$ is interpreted as $H(\vec a;z+i\varepsilon)$.

We conclude this section by giving an example of a sample program that prints all nine HPL's of weight two at the  point $z = 1.54+ 0.91i$.
\newpage
\begin{verbatim}
    program chaplintest
   
    double complex HPL2,z
    integer n1,n2

    z = dcmplx(1.54d0,0.91d0)
    do n1=-1,1
      do n2=-1,1
        print*, HPL2(n1,n2,z)
      enddo
    enddo

    end program
\end{verbatim}
\subsection{Validation and comparison to other codes}
We have compared the results obtained by {\sc Chaplin} against {\tt hplog} and GiNaC for real values of the argument and against HPL and GiNaC for complex values of the argument. The results for some sample points are summarized in Tab.~\ref{tab:z=0.5} - \ref{tab:z=2+2i}. For real arguments, a small imaginary part is assumed. Note that the reduction of $H(1,-1,-1,0;z)$ to basis functions is among the most complicated cases, involving almost 50 lines of {\tt Fortran} code.

In order to give an idea of the CPU time needed per function call, we present in Tab.~ \ref{tab:time} the time for one million function calls on a 3 GHz Intel X5450 Processor for some HPL's in the six different regimes of the complex plane shown in Fig.~\ref{fig}. The running time varies only marginally inside a given region. The resulting average times for a single function call are given in Tab.~\ref{tab:time} in units of microseconds ($\mu s$).
\begin{equation*}
\phantom{a}
\end{equation*}
\begin{equation*}
\phantom{a}
\end{equation*}

\begin{center}
\begin{table}[!ht]
\begin{center}
\begin{tabular}{|c|c|c|c|c|c|c|}
\hline
Region & I & II & III & IV & V & IV\\
\hline\hline
$\mathrm{Li}_2(z)$& 1.4 & 4.4 & 4.5 & 0.2 & 0.2 & 0.3\\
\hline
$\mathrm{Li}_3(z)$& 1.9 & 4.7 & 5.0 & 0.2 & 0.2 & 0.3\\
\hline
$\mathrm{Li}_4(z)$& 6.2 & 8.9 & 9.1 & 4.2 & 4.3 & 4.5\\
\hline
$H(0,1,0,-1;z)$& 8.7 & 17.3 & 44.2 & 4.3 & 4.4 & 4.5\\
\hline
$H(1,-1,-1,0;z)$& 140.0 & 213.4 & 393.8 & 4.3 & 4.8 & 4.7\\
\hline
\end{tabular}
\caption{Average time per function call in microseconds ($\mu s$), for the six regions of the complex plane shown in Fig.~\ref{fig}.}
\label{tab:time}
\end{center}
\end{table}
\end{center}
\begin{center}
\begin{table}[!th]
\begin{center}
 \begin{tabular}{|c|c|c|}
\hline
\multicolumn{3}{|c|}{$z = 0.5$}\\ 
\hline
& \sc Chaplin &   5.8224052646501201e-01       $+$    0.       $i$ \\ 
$\mathrm{Li}_2(z)$ & \tt hplog &   5.8224052646501256e-01       $+$    0.       $i$ \\ 
& GiNaC & 5.8224052646501245e-01 $+$ 0. $i$ \\ 
\hline
& \sc Chaplin &   5.3721319360804021e-01       $+$    0.       $i$ \\ 
$\mathrm{Li}_3(z)$ & \tt hplog &   5.3721319360804010e-01       $+$    0.       $i$ \\ 
& GiNaC & 5.3721319360804021e-01 $+$ 0. $i$ \\ 
\hline
& \sc Chaplin &   5.1747906167389901e-01       $+$    0.       $i$ \\ 
$\mathrm{Li}_4(z)$ & \tt hplog &   5.1747906167389945e-01       $+$    0.       $i$ \\ 
& GiNaC & 5.1747906167389934e-01 $+$ 0. $i$ \\ 
\hline
& \sc Chaplin &   7.7856141848313908e-02  $+$    0.       $i$ \\ 
$\mathrm{H}(0,1,0,-1;z)$ & \tt hplog &   7.7856141848313090e-02  $+$    0.       $i$ \\ 
& GiNaC & 7.7856141848313215e-02 $+$ 0. $i$ \\ 
\hline
& \sc Chaplin &  -6.3908284909225732e-02  $+$    0.     
 $i$ \\ 
$\mathrm{H}(1,-1,-1,0;z)$ & \tt hplog &  -6.3908284909226009e-02  $+$    0.     
 $i$ \\ 
& GiNaC & -6.3908284909226135e-02 $+$ 0.
 $i$ \\ 
\hline
\end{tabular}
\caption{\label{tab:z=0.5}Comparison between {\sc Chaplin}, {\tt hplog} and GiNaC for the point $z = 0.5$.}
\end{center}
\end{table}
\end{center}
\begin{center}
\begin{table}[!th]
\begin{center}
\begin{tabular}{|c|c|c|}
\hline
\multicolumn{3}{|c|}{$z = 2.0$}\\ 
\hline
& \sc Chaplin &    2.4674011002723404e+00       $+$    2.1775860903036022e+00       $i$ \\ 
$\mathrm{Li}_2(z)$ & \tt hplog &    2.4674011002723399e+00       $+$    2.1775860903036017e+00       $i$ \\ 
& GiNaC & 2.4674011002723395e+00 $-$ 2.1775860903036022e+00 $i$ \\ 
\hline
& \sc Chaplin &    2.7620719062289245e+00       $+$   7.5469382946024677e-01       $i$ \\ 
$\mathrm{Li}_3(z)$ & \tt hplog &    2.7620719062289241e+00       $+$   7.5469382946024799e-01       $i$ \\ 
& GiNaC & 2.7620719062289241e+00 $-$ 7.5469382946024810e-01 $i$ \\ 
\hline
& \sc Chaplin &    2.4278628067547032e+00       $+$   1.7437130002545320e-01       $i$ \\ 
$\mathrm{Li}_4(z)$ & \tt hplog &    2.4278628067547032e+00       $+$   1.7437130002545298e-01       $i$ \\ 
& GiNaC & 2.4278628067547032e+00 $-$ 1.7437130002545306e-01 $i$ \\ 
\hline
& \sc Chaplin &   5.1994752047739468e-01       $+$    1.7909927717176168e+00       $i$ \\ 
$\mathrm{H}(0,1,0,-1;z)$ & \tt hplog &   5.1994752047739512e-01       $+$    1.7909927717176164e+00       $i$ \\ 
& GiNaC & 5.1994752047739445e-01 $-$ 1.7909927717176168e+00 $i$ \\ 
\hline
& \sc Chaplin &   8.0548200591357266e-01       $-$   1.3189461296972327e+00     
 $i$ \\ 
$\mathrm{H}(1,-1,-1,0;z)$ & \tt hplog &   8.0548200591356789e-01       $-$   1.3189461296972333e+00     
 $i$ \\ 
& GiNaC & 8.0548200591356811e-01 $+$ 1.3189461296972318e+00
 $i$ \\ 
\hline
\end{tabular}
\caption{\label{tab:z=2}Comparison between {\sc Chaplin}, {\tt hplog} and GiNaC for the point $z = 2.0$. Note that GiNaC uses a different convention for the imaginary parts.}
\end{center}
\end{table}
\end{center}
 \begin{center}
\begin{table}[!th]
\begin{center}
\begin{tabular}{|c|c|c|}
\hline
\multicolumn{3}{|c|}{$z = 0.5 + 0.5\,i$}\\ 
\hline
& \sc Chaplin &   4.5398526915029508e-01       $+$   6.4376733288926902e-01       $i$ \\ 
$\mathrm{Li}_2(z)$ & HPL & 4.5398526915029541e-01 $+$ 6.4376733288926879e-01 $i$\\ 
& GiNaC & 4.5398526915029558e-01 $+$ 6.4376733288926880e-01 $i$ \\ 
\hline
& \sc Chaplin &   4.8615953708555987e-01       $+$   5.7007740708876864e-01       $i$ \\ 
$\mathrm{Li}_3(z)$ & HPL & 4.8615953708556014e-01 $+$ 5.7007740708876930e-01 $i$\\ 
& GiNaC & 4.8615953708556009e-01 $+$ 5.7007740708876897e-01 $i$ \\ 
\hline
& \sc Chaplin &   4.9578112182183881e-01       $+$   5.3402238407975344e-01       $i$ \\ 
$\mathrm{Li}_4(z)$ & HPL &4.9578112182183897e-01 $+$ 5.3402238407975377e-01 $i$ \\ 
& GiNaC & 4.9578112182183876e-01 $+$ 5.3402238407975355e-01 $i$ \\ 
\hline
& \sc Chaplin &  -3.6325772179994109e-02  $+$   1.384991682646747e-01       $i$ \\ 
$\mathrm{H}(0,1,0,-1;z)$ & HPL & -3.6325772179994879e-02 $+$ 1.3849916826467456e-01 $i$\\ 
& GiNaC & -3.6325772179994845e-02 $+$ 1.3849916826467457e-01 $i$ \\ 
\hline
& \sc Chaplin &   9.1142643382278842e-02  $-$  9.8191320890700317e-02
 $i$ \\ 
$\mathrm{H}(1,-1,-1,0;z)$ & HPL &  9.1142643382278176e-02 $-$ 9.8191320890700678e-02 $i$\\ 
& GiNaC & 9.1142643382278163e-02 $-$ 9.8191320890700595e-02
 $i$ \\ 
\hline
\end{tabular}
\caption{\label{tab:z=0.5+0.5i}Comparison between {\sc Chaplin}, HPL and GiNaC for the point $z = 0.5 + 0.5\,i$.}
\end{center}
\end{table}
\end{center}
\begin{center}
\begin{table}[!th]
\begin{center}
\begin{tabular}{|c|c|c|}
\hline
\multicolumn{3}{|c|}{$z = 2.0 + 2.0\, i$}\\ 
\hline
& \sc Chaplin &   3.4497312626178323e-01       $+$    2.7342872186403562e+00       $i$ \\ 
$\mathrm{Li}_2(z)$ & HPL & 3.4497312626178264e-01 $+$ 2.7342872186403560e+00 $i$ \\ 
& GiNaC & 3.4497312626178261e-01 $+$ 2.7342872186403562e+00 $i$ \\ 
\hline
& \sc Chaplin &    1.2370548907501702e+00       $+$    2.7024607822310069e+00       $i$ \\ 
$\mathrm{Li}_3(z)$ & HPL & 1.2370548907501697e+00 $+$ 2.7024607822310064e+00 $i$\\ 
& GiNaC & 1.2370548907501697e+00 $+$ 2.7024607822310065e+00 $i$ \\ 
\hline
& \sc Chaplin &    1.7008027579027265e+00       $+$    2.4625762177390942e+00       $i$ \\ 
$\mathrm{Li}_4(z)$ & HPL & 1.7008027579027260e+00 $+$ 2.4625762177390939e+00 $i$ \\ 
& GiNaC & 1.7008027579027261e+00 $+$ 2.4625762177390937e+00 $i$ \\ 
\hline
& \sc Chaplin &   -1.3092921033357463e+00       $+$   8.6009513536901472e-01       $i$ \\ 
$\mathrm{H}(0,1,0,-1;z)$ & HPL & -1.3092921033357458e+00 $+$ 8.6009513536901561e-01 $i$\\ 
& GiNaC & -1.3092921033357459e+00 $+$ 8.6009513536901561e-01 $i$ \\ 
\hline
& \sc Chaplin &    1.3154184588794104e+00       $-$ 2.6274818437873471e-01     
 $i$ \\ 
$\mathrm{H}(1,-1,-1,0;z)$ & HPL & 1.3154184588794054e+00 $-$ 2.6274818437872689e-01 $i$ \\ 
& GiNaC & 1.3154184588794056e+00 $-$ 2.6274818437872688e-01
 $i$ \\ 
\hline
\end{tabular}
\caption{\label{tab:z=2+2i}Comparison between {\sc Chaplin}, HPL and GiNaC for the point $z = 2.0 + 2.0\,i$.}
\end{center}
\end{table}
\end{center}
\section{Summary}
In this paper we have presented {\sc Chaplin}, a new {\tt Fortran} library to compute harmonic polylogarithms up to weight four for arbitrary complex argument. The algorithm is based on a reduction of HPL's to a set of basis functions which are then evaluated numerically using series expansions allowing for a very fast numerical convergence, hence rendering the computational cost of a function call quite modest. 
We have checked our numerical results against well-established codes~\cite{Gehrmann:2001pz, Maitre:2005uu, Maitre:2007kp, Vollinga:2004sn} and found agreement to at least 14 digits for any argument in the complex plane.
\enlargethispage{5mm}
\section*{Acknowledgements}
The authors are grateful to B.~Anastasiou, H.~Gangl, T.~Gehrmann and J.~Rhodes for valuable discussions and comments. This work was supported by the Research Executive Agency (REA) of the European Union under the Grant Agreement number PITN-GA-2010-264564 (LHCPhenoNet) and by the Swiss National Foundation
under contract SNF 200020-126632.

\begin{equation*}\bsp
\phantom{a}&\phantom{a}\\
\phantom{a}&\phantom{a}\\
\phantom{a}&\phantom{a}\\
\phantom{a}&\phantom{a}\\
\phantom{a}&\phantom{a}\\
\phantom{a}&\phantom{a}\\
\phantom{a}&\phantom{a}\\
\phantom{a}&\phantom{a}\\
\esp\end{equation*}


\appendix

\section{Proof of the associativity of the convolution product}
\label{app:Associativity}
In this appendix we proof the associativity of the convolution product, Eq.~\eqref{eq:convol_props}. Using the definition of the convolution product, Eq.~\eqref{eq:convolution_def}, we obtain,
\beq\bsp
((a\ast b)\ast c)_N =\sum_{m=0}^N\binom{N}{m}(a\ast b)_m\,c_{N-m} = \sum_{m=0}^N\sum_{n=0}^m \binom{N}{m}\binom{m}{n}\,a_n\,b_{m-n}\,c_{N-m}\,.
\esp\eeq
We now exchange the sums over $m$ and $n$ and shift the summation variable such that all sums start from zero. This gives,
\beq\bsp
((a\ast b)\ast c)_N &\,=\sum_{n=0}^N\sum_{m=n}^N \binom{N}{m}\binom{m}{n}\,a_n\,b_{m-n}\,c_{N-m}\\
&\,=\sum_{n=0}^N\sum_{m=0}^{N-n} \binom{N}{m+n}\binom{m+n}{n}\,a_n\,b_{m}\,c_{N-m-n}\\
&\,=\sum_{n=0}^N\sum_{m=0}^{n} \binom{N}{m+N-n}\binom{m+N-n}{N-n}\,a_{N-n}\,b_{m}\,c_{n-m}\,,
\esp\eeq
where the last step follows from changing the summation variable according to $n\to N-n$. The product of binomials can be simplified,
\beq\bsp
\binom{N}{m+N-n}\binom{m+N-n}{N-n} &\,= {N!\over (N+m-n)!(n-m)!}\,{(N+m-n)!\over m!(N-n)!}\\ 
&\,= {N!\over n!(N-n)}\,{n!\over m!(n-m)!} = \binom{N}{n}\binom{n}{m}\,,
\esp\eeq
yielding,
\beq
((a\ast b)\ast c)_N = \sum_{n=0}^N\sum_{m=0}^{n} \binom{N}{n}\binom{n}{m}\,a_{N-n}\,b_{m}\,c_{n-m} = (a\ast (b\ast c))_N\,.
\eeq

\section{Derivation of Eq.~\eqref{eq:HPL_Logx}}
\label{app:proofs}
In this appendix we present the derivation of the series expansions given in Eq.~\eqref{eq:HPL_Logx}. The derivations follow the same spirit as in all other cases discussed in Section~\ref{sec:series}, \emph{i.e.}, we start from the integral representation and perform a change of variable, before inserting the Taylor expansions of the integrand. The proof involves some technical issues, which are discussed in detail in the following on the example of $H(0,1,0,-1;e^x)$. All other cases are similar.

To derive the Taylor expansion of $H(0,1,0,-1;e^x)$, we start from the series expansion of
$H(-1;e^x) = \log(1+e^x)$ and then integrate up to $H(0,1,0,-1;e^x)$. 
The series expansion of $H(-1;e^x) = \log(1+e^x)$ is easily obtained from the integral representation and the generating function of the Genocchi numbers, Eq.~\eqref{eq:Gen_Func_Genocchi}. Letting $t=e^{t'}$, we get
\beq\bsp
H(-1;e^x) &\,= H(-1;1) + \int_1^{e^x}{\rd t\over 1+t}=\log2 - {1\over2}\,\int_0^x{\rd t'\over t'}\,{2\,(-t')\over 1+e^{-t'}}\\
&\,=\log2 - {1\over2}\,\sum_{n=0}^\infty{\overline{G}_n\over n!}\,\int_0^x\rd t'\,t'^{n-1}=\log2-{1\over2}\,\sum_{n=0}^\infty{\overline{G}_n\over n!}\,{x^n\over n}\\
&\,=\log2-{1\over2}\,\sum_{n=0}^\infty{\gamma_n\over n!}\,x^n\,,
\esp\eeq
where we introduced the shorthand
\beq
\gamma_n={\overline G_n\over n}\,.
\eeq

Next we determine the series expansion of $H(0,-1;e^x)=-\li{2}(-e^x)$. Repeating exactly the same steps as for $H(-1;e^x)$, we obtain
\beq\bsp
H(0,-1;e^x) &\,= -\li{2}(-1) + \int_1^{e^x}{\rd t\over t}\,H(-1;t) = {\pi^2\over 12}+\int_0^x\rd t'\,H(-1;e^{t'})\\
&\,={\pi^2\over 12} + \log2\,x-{1\over2}\,\sum_{n=0}^\infty{\gamma_n\over (n+1)!}\,x^{n+1}\,.
\esp\eeq

If we try to repeat the same procedure for $H(1,0,-1;e^{x})$, a technical difficulty arises: in the previous example we had the split the integral over $[0,e^x]$ into two contributions from $[0,1]$ and $[1,e^x]$. In the present case, however, we cannot do this, because $H(1,0,-1;1)$ is divergent. We therefore introduce a regulator $\varepsilon$ and split the integration region into $[0,e^\varepsilon]$ and $[e^\varepsilon,e^x]$, and we will take the limit $\varepsilon\to0$ at the very end. We then obtain,
\beq\bsp
H(1,0,-1;e^x) &\,= H(1,0,-1;e^\varepsilon) + \int_{e^\varepsilon}^{e^x}{\rd t\over 1-t}\,H(0,-1;t)\\
&\,=
H(1,0,-1;e^\varepsilon) - \int_{\varepsilon}^{x}{\rd t'\over t'}{(-t')\over e^{-t'}-1}\,H(0,-1;e^{t'})\\
&\,= H(1,0,-1;e^\varepsilon) +{\pi^2\over 12}\,[H(1;e^x) - H(1;e^\varepsilon)]\\
&\, + \log2\,[H(1,0;e^x) - H(1,0;e^\varepsilon)] +{1\over2}\,\sum_{n=0}^\infty{(\overline{B}\ast\mathring{\gamma})_n\over (n+1)!}\,x^{n+1}\,.
\esp\eeq
We now have to send the regulator to zero. Concentrating only on terms depending on $\varepsilon$ and using the shuffle algebra for HPL's, we obtain,
\beq\bsp
H&(1,0,-1;e^\varepsilon) -{\pi^2\over 12}\,H(1;e^\varepsilon)-\log2\,H(1,0;e^\varepsilon)\\
&=H(1;e^\varepsilon)\,H(0,-1;e^\varepsilon) - {\pi^2\over 12}\,H(1;e^\varepsilon)\\
& - H(0,1,-1;e^\varepsilon)- H(0,-1,1;e^\varepsilon)-\log2\, H(1,0;e^\varepsilon)\,.
\esp\eeq
All the terms in the second line have a smooth limit as $\varepsilon\to0$, and the two divergent terms in the first line cancel exactly. This leaves us with
\beq\bsp
H(1,0,-1;e^x) &\,= -\frac{5}{8}\,\zeta_3-\frac{\pi^2}{12} \,\log 2+{\pi^2\over 12}\,H(1;e^x) + \log2\, H(1,0;e^x)\\&\,+{1\over2}\,\sum_{n=0}^\infty{(\overline{B}\ast\mathring{\gamma})_n\over (n+1)!}\,x^{n+1}\,.
\esp\eeq

The last integration is easy to perform and we immediately obtain
\beq\bsp
H&(0,1,0,-1;e^x) =-4\, \text{Li}_4\left(\frac{1}{2}\right)-\frac{5}{2}\, \zeta_3\, \log2+\frac{17 \pi ^4}{480}-\frac{1}{6}\,\log ^42\\
&+\frac{\pi ^2 }{6}\, \log ^22-{5\over8}\,\zeta_3\,x+{\pi^2\over 6}\,\log2\,x
+{\pi^2\over 12}\,\li{2}(e^x)\\
&+\log2\,x\,\li{2}(e^x)-2\log2\,\li{3}(e^x)
+{1\over 2}\sum_{n=0}^\infty{(\overline{B}\ast\mathring{\gamma})_n\over(n+2)!}\,x^{n+2}\,,
\esp\eeq
which agrees with Eq.~\eqref{eq:HPL_Logx} after replacing $x=\log z$.

\end{document}